\documentclass{PoS}

\usepackage{graphicx}
\usepackage{amsmath}
\usepackage{amssymb}
\usepackage{longtable}

\newcommand{\be}{\begin{equation}}
\newcommand{\ee}{\end{equation}}
\newcommand{\bea}{\begin{eqnarray}}
\newcommand{\eea}{\end{eqnarray}}
\newcommand{\bi}{\begin{itemize}}
\newcommand{\ei}{\end{itemize}}
\newcommand{\ben}{\begin{enumerate}}
\newcommand{\een}{\end{enumerate}}
\newcommand{\bt}{\begin{tabbing}}
\newcommand{\et}{\end{tabbing}}
\newcommand{\nn}{\nonumber}

\newcommand{\calO}{{\mathcal O}}

\newcommand{\bfp}{{\bf p}}
\newcommand{\bfq}{{\bf q}}

\newcommand{\bfx}{{\bf x}}

\newcommand{\crad}{\langle r^2 \rangle}

\title{
   \begin{picture}(0,0)(0,0)%
   \put(355,75){\makebox(0,0)[l]{\textnormal{\normalsize KEK-CP-227}}}%
   \end{picture}%
   Pion form factors from lattice QCD with exact chiral symmetry
}

\ShortTitle{Pion form factors from lattice QCD with exact chiral symmetry}

\author{
   Takashi Kaneko\thanks{E-mail: takashi.kaneko@kek.jp} 
   \\
   \\
   \\
   KEK Theory Center, 
   High Energy Accelerator Research Organization (KEK),
   Ibaraki 305-0801, Japan 
   \\
   School of High Energy Accelerator Science,
   The Graduate University for Advanced Studies (Sokendai),
   Ibaraki 305-0801, Japan
}

\abstract{
We calculate pion vector and scalar form factors 
in two-flavor lattice QCD and study the chiral behavior 
of the vector and scalar radii $\crad_{V,S}$.
For a direct comparison with chiral perturbation theory (ChPT),
chiral symmetry is exactly preserved by employing the overlap quark action.
We utilize the all-to-all quark propagator
in order to calculate
the scalar form factor including the contributions of disconnected diagrams.
A detailed comparison with ChPT reveals that 
two-loop contributions are important to describe 
the chiral behavior of the radii
in our region of the pion mass $M_\pi \! \gtrsim 290$~MeV. 
From chiral extrapolation based on two-loop ChPT, 
we obtain 
$\crad_V \!=\! 0.409(23)(37)~\mbox{fm}^2$ and 
$\crad_S \!=\! 0.617(79)(66)~\mbox{fm}^2$, 
which are consistent with phenomenological analyses.
}

\FullConference{6th International Workshop on Chiral Dynamics, CD09\\
		July 6-10, 2009\\
		Bern, Switzerland}

\begin{document}


\section{Introduction}
\vspace{-3mm}


A detailed study of the chiral behavior of pion observables
is an important subject
towards a deep understanding of the low-energy dynamics of QCD.
The pion vector form factor $F_V(q^2)$ 
has been precisely measured by experiments and 
the charge radius $\crad_V$ can be extracted in a model-independent way.
Although a detailed comparison of $\crad_V$ 
between chiral perturbation theory (ChPT)
and non-perturbative determinations in lattice QCD
is very interesting,
the chiral behavior of $\crad_V$ is distorted 
by explicit chiral symmetry breaking
with the conventional lattice quark actions.
 
The chiral behavior of the scalar form factor $F_S(q^2)$ 
is another interesting subject.
The scalar radius $\crad_S$ provides a determination of 
the low-energy constant (LEC) $l_4$,
which should be compared with that from $F_\pi$, 
and $\crad_S$ has 6~times enhanced chiral logarithm compared to $\crad_V$.
Since there are no experimental processes directly related to $F_S(q^2)$,
its direct determination is possible only through lattice QCD.
It is, however, computationally very demanding 
to evaluate contributions of disconnected diagrams to $F_S(q^2)$ 
with the conventional simulation methods.

In this article,
we present our calculation of these pion form factors 
in two-flavor lattice QCD.
Chiral symmetry is exactly preserved by employing 
the overlap quark action \cite{Overlap:NN}
for a direct comparison of $\crad_{V,S}$ with ChPT at two loops.
We use the so-called all-to-all quark propagator \cite{A2A} 
to calculate $F_S(q^2)$ including the disconnected contributions.
We refer readers to Ref.~\cite{PFF:Nf2:RG+Ovr:JLQCD}
for more detailed description of this study.


\vspace{-3mm}
\section{Simulation method}
\vspace{-3mm}


We simulate QCD with two flavors of degenerate up and down quarks
using the Iwasaki gauge action and the overlap quark action. 
We also introduce a topology fixing term \cite{exW+extmW:JLQCD}
into the lattice action to substantially reduce the computational cost.
Our numerical simulations are carried out 
on a $N_s^3 \times N_t \!=\! 16^3 \times 32$ lattice 
at a lattice spacing of $a\!=\!0.1184(21)$~fm,
which is fixed from the heavy quark potential.
In the trivial topological sector $Q\!=\!0$, 
we simulate four values of the bare up and down quark masses 
$m_{ud}\!=\!0.015,0.025,0.035$ and 0.050, 
which cover a range of the pion mass 
$290 \! \lesssim \! M_\pi[\mbox{MeV}] \! \lesssim \! 520$.
Statistics are 100 independent configurations at each $m_{ud}$.
We also simulate nontrivial topological sectors $Q\!=\!-2$ and $-4$
to study effects of the fixed topology.
Further details on our configuration generation
are presented in Ref.\cite{Prod_Run:JLQCD:Nf2:RG+Ovr}.

\begin{figure}[t]
\begin{center}
   \includegraphics[angle=0,width=0.3\linewidth,clip]%
                   {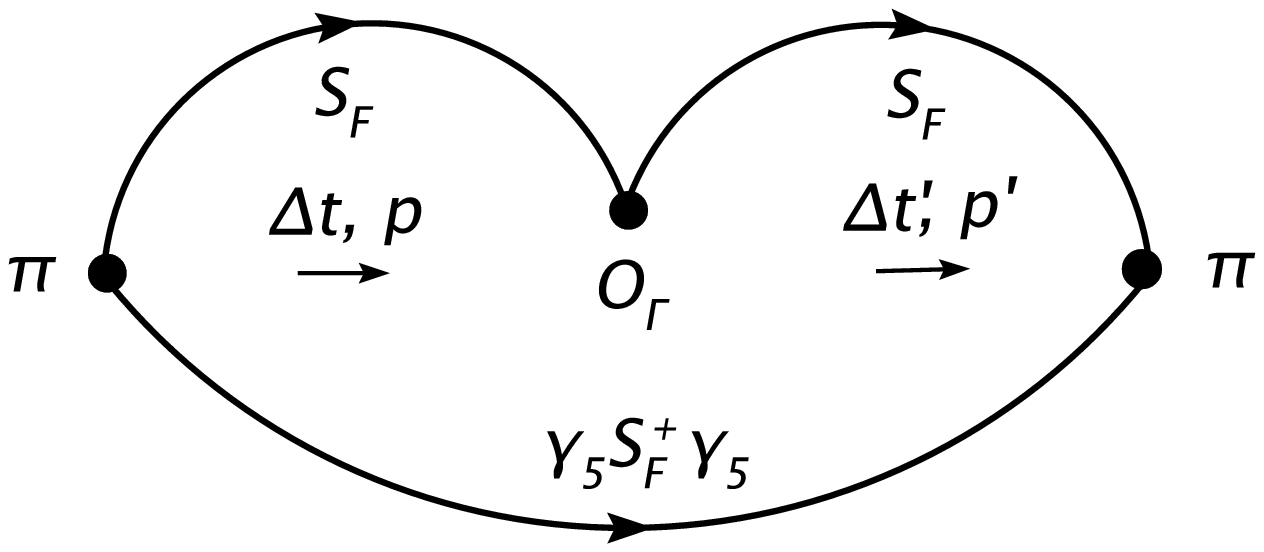}
   \hspace{5mm}
   \includegraphics[angle=0,width=0.3\linewidth,clip]%
                   {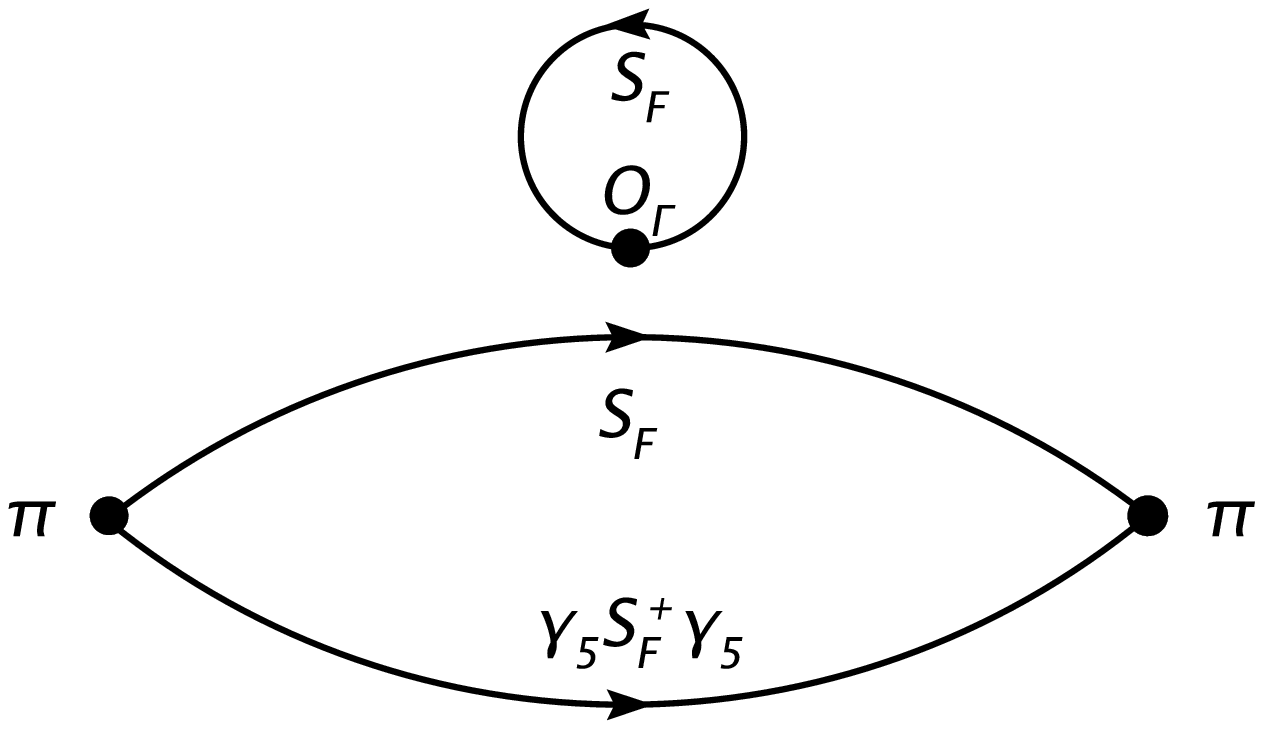}
   \hspace{5mm}
   \includegraphics[angle=0,width=0.3\linewidth,clip]%
                   {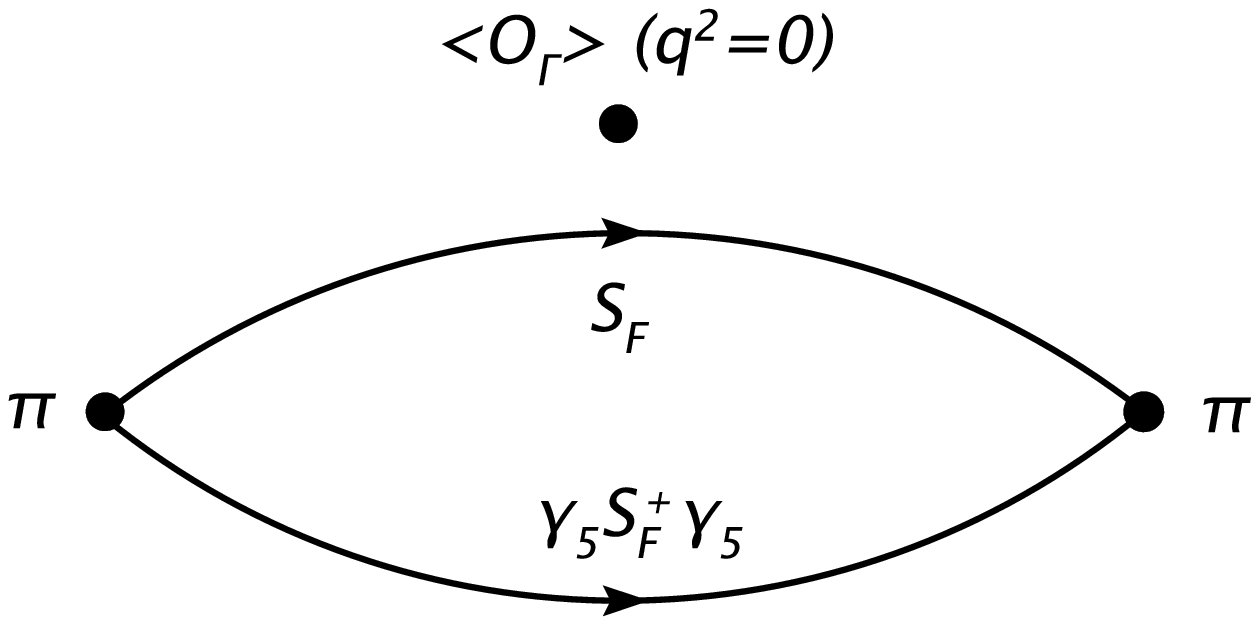}
   \vspace{-2mm}
   \caption{
      Connected (left-most diagram) and disconnected (middle diagram) 
      three point functions.
      The right-most diagram with the vacuum expectation value (VEV)
      $\langle \calO_\Gamma \rangle$ 
      contributes to $F_S(0)$. 
      Each line shows quark propagator $S_F$.
      We use $S_F(x,x^\prime)\!=\!\gamma_5 S_F(x^\prime,x)^\dagger \gamma_5$
      for the spectator quark.
   }
   \label{fig:meas:corr_3pt:diag}
   \vspace{-7mm}
\end{center}
\end{figure}

The matrix element
$\langle \pi(p^\prime) | \calO_\Gamma(p^\prime\!-\!p) | \pi(p) \rangle$
can be extracted from the three-point function
\bea
   C_{\pi \calO_\Gamma \pi}(\Delta t, \Delta t^\prime; \bfp, \bfp^\prime)
   & = &
   \frac{1}{N_s^3 N_t}
   \sum_{\bfx,t}
   C_{\pi \calO_\Gamma \pi}(\bfx, t; \Delta t, \Delta t^\prime; \bfp, \bfp^\prime),
   \label{eqn:meas:corr_3pt}
   \\
   C_{\pi \calO_\Gamma \pi}(\bfx, t; \Delta t, \Delta t^\prime; \bfp, \bfp^\prime)
   & = &
   \sum_{\bfx^\prime \bfx^{\prime\prime}} 
      \langle 
         \calO_\pi(\bfx^\prime,t+\Delta t+\Delta t^\prime)   
         \calO_\Gamma (\bfx^{\prime\prime},t+\Delta t)
         \calO_\pi(\bfx,t)^\dagger   
         e^{-i\bfp^\prime (\bfx^\prime-\bfx^{\prime\prime})}
         e^{-i\bfp (\bfx^{\prime\prime}-\bfx)}
      \rangle,
   \hspace{8mm}
   \label{eqn:meas:corr_3pt:piece}
\eea
where 
$\calO_\Gamma$ is the vector current $V_\mu$ or scalar operator $S$,
and $\calO_\pi^\dagger$ represents 
an interpolating field to create the physical pion state.
In the conventional method,
one calculates the so-called point-to-all quark propagator $S_F(x^\prime,x)$,
which flows from a fixed lattice site $x$ to any site $x^\prime$,
by solving
\bea
   \sum_{x^\prime}D(y,x^\prime)S_F(x^\prime,x)
   & =&
   \delta_{y,x},
   \label{eqn:meas:lineq}
\eea
where $D$ is the Dirac operator.
Then, $C_{\pi \calO_\Gamma \pi}(\bfx, t; \Delta t, \Delta t^\prime; \bfp, \bfp^\prime)$
can be calculated by connecting the point-to-all propagators 
as shown in Fig.~\ref{fig:meas:corr_3pt:diag}.
However, 
we need to solve Eq.~(\ref{eqn:meas:lineq}) for each lattice site $x$
to evaluate the disconnected quark loop $D^{-1}(x,x)$
as well as to average 
$C_{\pi \calO_\Gamma \pi}(\bfx, t; \Delta t, \Delta t^\prime; \bfp, \bfp^\prime)$
over the pion source location
$x\!=\!(\bfx,t)$ as in Eq.~(\ref{eqn:meas:corr_3pt}).
This needs prohibitively large CPU cost.

This difficulty can be avoided by constructing the all-to-all quark propagator,
which contains the quark propagating from any lattice site to any site,
in an effective way.
Along the strategy proposed in Ref.~\cite{A2A},
we prepare 100 low-lying modes of $D$ for each gauge configuration,
and their contribution to the all-to-all quark propagator is 
calculated exactly as 
\bea
   (S_F)_{\rm low}(x,y) 
   & = &
   \sum_{k=1}^{100}
   \frac{1}{\lambda_k} u_k(x) u_k^\dagger(y),
   \label{meas:a2a:prop:low}
\eea
where $\lambda_k$ and $u_k$ represent $k$-th smallest eigenvalue 
and corresponding eigenvector.
It is expected that this low-mode contribution dominates 
low-energy dynamics of pions.
Possibly small contribution from the remaining high-modes
can be estimated stochastically by the so-called noise method \cite{noise}, 
which is not computationally intensive.
We refer to Ref.~\cite{PFF:Nf2:RG+Ovr:JLQCD}
for more technical details on our method 
to calculate pion correlators using the all-to-all propagator.

We calculate three-point functions, $C_{\pi V_4 \pi}$ and $C_{\pi S \pi}$, 
as well as the two-point function 
\bea
   C_{\pi \pi}(\Delta t; \bfp)
   & = &
   \frac{1}{N_s^3 N_t}
   \sum_{\bfx^\prime} \sum_{\bfx,t}
      \langle 
         \calO_\pi(\bfx^\prime,t+\Delta t)   
         \calO_\pi(\bfx,t)^\dagger   
         e^{-i\bfp (\bfx^\prime-\bfx)}
      \rangle.
   \label{eqn:meas:corr_2pt}
\eea
We take 33 choices for the spatial momentum $\bfp^{(\prime)}$ 
with $|\bfp^{(\prime)}|\!\leq\!2$,
which cover a region of the momentum transfer 
$-1.7 \! \lesssim q^2~\mbox{[GeV$^2$]} \! \leq \! 0$
for the three-point functions.
Note that the spatial momentum is shown in units of $2\pi/L$ 
in this article.

\section{Determination of pion form factors}

\begin{figure}[t]
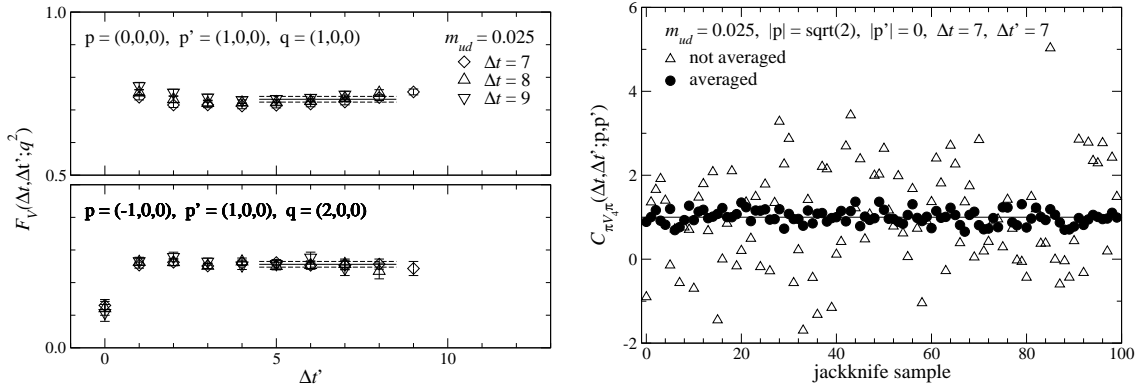

\begin{center}
   \includegraphics[angle=0,width=0.48\linewidth,clip]%
                   {./pff_v_vs_dtsnk_m0025.eps}
   \hspace{3mm}
   \includegraphics[angle=0,width=0.48\linewidth,clip]%
                   {./jkd_msn_3pt_p-p-v_mud1_mval111_mom0700_smr11.eps}

   \vspace{-2mm}
   \caption{
      Left panel : 
      effective value $F_V(\Delta t, \Delta t^\prime;q^2)$ 
      at $m_{ud}\!=\!0.025$,
      which is around a quarter of the physical strange quark mass.
      Right panel : 
      statistical fluctuation of 
      $C_{\pi V_4 \pi}(\Delta t, \Delta t^{\prime};\bfp,\bfp^\prime)$ 
      at $(|\bfp|,|\bfp^\prime|)\!=\!(\sqrt{2},0)$ and 
      $\Delta t\!=\!\Delta t^\prime\!=\!7$.
      Filled and open symbols are results with and without 
      averaging over the source location $(\bfx,t)$.
   }
   \label{fig:pff:pff_v}
   \vspace{-5mm}
\end{center}
\end{figure}

For a precise lattice calculation of the form factors,
it is advantageous to construct an appropriate ratio of correlators
so that some uncertainties, such as renormalization factors,
cancel at least partially in the ratio \cite{dble_ratio}.
In this study, 
we calculate effective value of the vector form factor from 
\bea
   F_V(\Delta t,\Delta t^\prime;q^2)
   & = & 
   \frac{2\,M_\pi}{E_\pi(|\bfp|)+E_\pi(|\bfp^\prime|)}
   \frac{R_V(\Delta t,\Delta t^\prime; |\bfp|,|\bfp^\prime|,q^2)}
        {R_V(\Delta t,\Delta t^\prime; 0,0,0)},
   \label{eqn:pff:dratio:pff_v}
   \\
   R_V(\Delta t,\Delta t^\prime; |\bfp|,|\bfp^\prime|,q^2)
   & = &
   \frac{1}{N_{|\bfp|,|\bfp^\prime|}}
   \sum_{\mbox{\scriptsize fixed } |\bfp|,|\bfp^\prime|}
   \frac{C_{\pi V_4 \pi}(\Delta t,\Delta t^\prime; \bfp,\bfp^\prime)}
        {C_{\pi \pi}(\Delta t;\bfp)\,
         C_{\pi \pi}(\Delta t^\prime;\bfp^\prime)},
   \label{eqn:pff:ratio:pff_v}
\eea       
where $(1/N_{|\bfp|,|\bfp^\prime|})\sum_{\mbox{\scriptsize fixed } |\bfp|,|\bfp^\prime|}$
represents the average over momentum configurations $\{(\bfp,\bfp^\prime)\}$
corresponding to the given value of $q^2$.
As shown in the left panel of Fig.~\ref{fig:pff:pff_v},
we observe a very clear signal of $F_V(\Delta t,\Delta t^\prime;q^2)$
with the statistical accuracy of typically 3\,--\,5\,\%.    
We note that 
the use of the all-to-all propagator 
enables us to achieve this high accuracy by averaging pion correlators 
over the location of the source operator 
$(\bfx,t)$ in Eqs.~(\ref{eqn:meas:corr_3pt}) and (\ref{eqn:meas:corr_2pt}).
The right panel of Fig.~\ref{fig:pff:pff_v} demonstrates 
the remarkable reduction of the statistical fluctuation of $C_{\pi V_4 \pi}$ 
by this averaging.

We determine the vector form factor $F_V(q^2)$ 
by a constant fit to the effective value
$F_V(\Delta t,\Delta t^\prime;q^2)$,
and include the correction due to the finite lattice volume 
estimated in one-loop ChPT \cite{FVC:PFF:V}.
Although 
we do not observe any significant $Q$ dependence of $F_V(q^2)$,
the spread in $F_V(q^2)$ among $Q\!=\!0$, $-2$, and $-4$ 
is taken as a conservative estimate of 
the systematic uncertainty due to the fixed topology.

\begin{figure}[b]
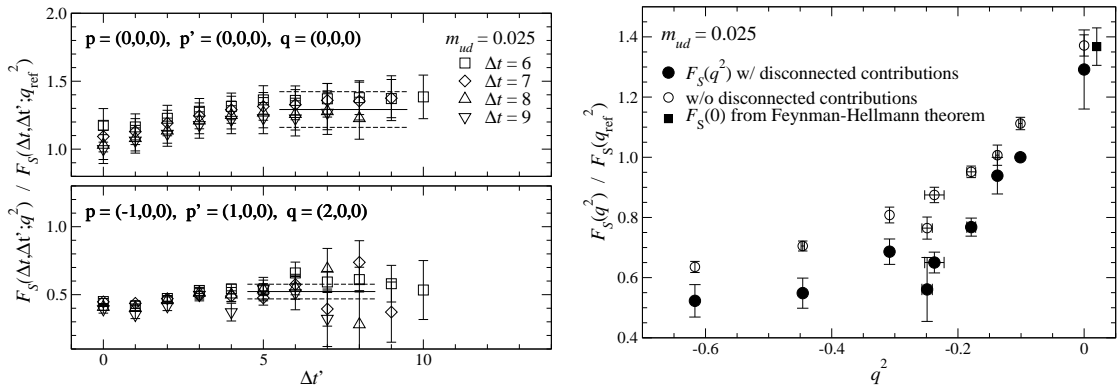

\begin{center}
   \includegraphics[angle=0,width=0.48\linewidth,clip]%
                   {./pff_s_vs_dtsnk_m0025.eps}
   \hspace{2mm}
   \includegraphics[angle=0,width=0.47\linewidth,clip]%
                   {./pff_s_vs_q2_norm-mom0100.eps}
   \vspace{-2mm}
   \caption{
      Left panel: 
      effective value of normalized scalar form factor 
      $F_S(\Delta t, \Delta t^\prime;q^2)/F_S(\Delta t, \Delta t^\prime;q_{\rm ref}^2)$ at $m_{ud}\!=\!0.025$.
      Right panel: 
      scalar form factor $F_S(q^2)$ with (filled symbols) 
      and without (open symbols) 
      the contributions of the disconnected diagrams .
      Both data are normalized by a common value $F_S(q_{\rm ref}^2)$
      including the disconnected contribution.
   }
   \vspace{-4mm}
   \label{fig:pff:pff_s}
\end{center}
\end{figure}

The scalar form factor normalized at a certain momentum transfer 
$q_{\rm ref}^2$ can be calculated from 
\vspace{-1mm}
\bea
   \frac{F_S(\Delta t,\Delta t^\prime;q^2)}
        {F_S(\Delta t,\Delta t^\prime;q_{\rm ref}^2)}
   & = &
   \frac{R_S(\Delta t,\Delta t^\prime; q^2)}
        {R_S(\Delta t,\Delta t^\prime; q_{\rm ref}^2)},
   \label{eqn:pff:dratio:pff_s}
   \\
   R_S(\Delta t,\Delta t^\prime; q^2)
   & = &
   \frac{1}{N_{|\bfp|,|\bfp^\prime|}}
   \sum_{\mbox{\scriptsize fixed } |\bfp|,|\bfp^\prime|}
   \frac{C_{\pi S \pi}(\Delta t,\Delta t^\prime; \bfp,\bfp^\prime)}
        {C_{\pi \pi}(\Delta t;\bfp)\,
         C_{\pi \pi}(\Delta t^\prime;\bfp^\prime)}.
   \label{eqn:pff:ratio:pff_s}
\eea
At $q^2\!=\!0$,
$C_{\pi S \pi}$ has an additional contribution
shown in Fig.~\ref{fig:meas:corr_3pt:diag}
due to the VEV of the scalar operator $S$.
The subtraction of this contribution leads to 
a relatively large uncertainty in $F_S(\Delta t,\Delta t^\prime; q^2)$ 
at $q^2\!=\!0$ 
than at $q^2\!\ne\!0$, as seen in the left of Fig.~\ref{fig:pff:pff_s}.
Although
the Feynman-Hellmann theorem $2 F_S(0)\!=\!\partial M_\pi^2 / \partial m_{ud}$ 
provides a better determination of $F_S(0)$,
it is subject to systematic uncertainties of 
the chiral extrapolation of $M_\pi^2$.
In our simulation setup, 
$F_S(q^2)$ has the smallest relative error 
at the smallest nonzero value of $|q_{\rm ref}^2|$
with $|\bfq_{\rm ref}|\!=\!1$.
We therefore use $F_S(q^2)$ normalized at this $q_{\rm ref}^2$ 
in the following analysis.

We determine $F_S(q^2)/F_S(q_{\rm ref}^2)$ 
from the effective value 
$F_S(\Delta t, \Delta t^\prime;q^2)/F_S(\Delta t, \Delta t^\prime;q_{\rm ref}^2)$
in a similar way to $F_V(q^2)$.
The right panel of Fig.~\ref{fig:pff:pff_s} 
compares $F_S(q^2)$ to that without the contribution of the disconnected diagrams.
We observe a significant deviation between the two data,
which implies  the importance of the disconnected contributions 
in a precision study of $F_S(q^2)$.


\section{Parametrization of $q^2$ dependence}

\begin{figure}[b]
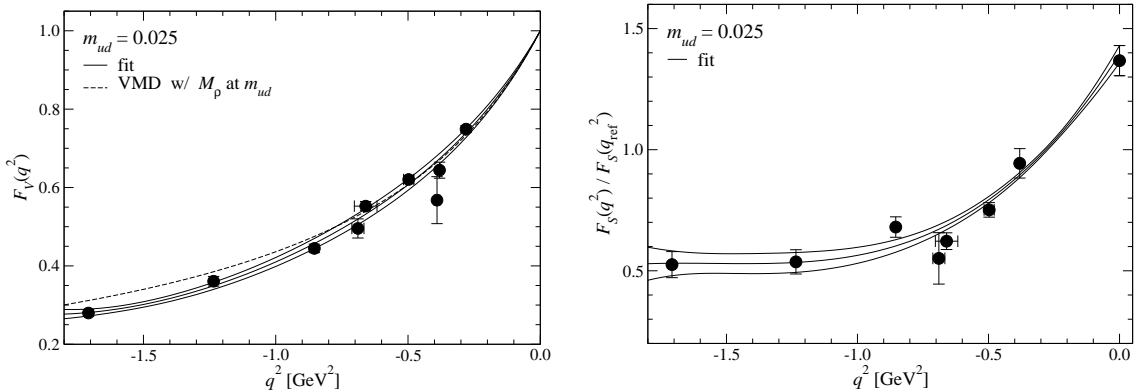

\begin{center}
   \includegraphics[angle=0,width=0.48\linewidth,clip]{./pff_v4_vs_q2_m0025.phys.eps}
   \hspace{3mm}
   \includegraphics[angle=0,width=0.48\linewidth,clip]{./pff_s_vs_q2_m0025.phys.eps}

   \vspace{-2mm}
   \caption{
      Vector form factor $F_V(q^2)$ (left panel) 
      and normalized scalar form factor $F_S(q^2)/F_S(q_{\rm ref}^2)$ 
      (right panel) as a function of $q^2$. 
      Solid lines show the fit curve and its error. 
      We also plot the $q^2$ dependence of $F_V(q^2)$ 
      expected from the VMD model by the dashed line.
   }
   \vspace{-5mm}
   \label{fig:q2:pff_v+pff_s}
\end{center}
\end{figure}

In Fig.~\ref{fig:q2:pff_v+pff_s},
we plot the vector form factor $F_V(q^2)$ 
and normalized scalar form factor $F_S(q^2)/F_S(q_{\rm ref}^2)$ 
as a function of $q^2$.
We observe that $F_V(q^2)$ is close to 
the vector meson dominance (VMD) hypothesis
$1/(1-q^2/M_\rho^2)$
with the vector meson mass $M_\rho$ measured at simulated $m_{ud}$.
%
%
We then assume that the small deviation due to the higher poles or cuts
can be approximated by a polynomial of $q^2$.
The $q^2$ dependence of $F_V(q^2)$ is therefore parametrized as 
\bea
   F_V(q^2)
   & = & 
   \frac{1}{1-q^2/M_{\rho}^2} + a_{V,1}\,q^2 
                              + a_{V,2}\,(q^2)^2 
                              + a_{V,3}\,(q^2)^3
   \hspace{1mm} = \hspace{1mm}
   1 + \frac{1}{6} \crad_V q^2 + c_V (q^2)^2 + \cdots
   \hspace{5mm}
   \label{eqn:q2:vs_q2:pff_v}
\eea
in order to determine the charge radius $\crad_V$ and the curvature $c_V$.
This form describes our data well as shown in Fig.~\ref{fig:q2:pff_v+pff_s}.
Results for $\crad_V$ and $c_V$ do not change significantly 
if we remove the cubic term or if we add higher order terms
into the parametrization form.

Due to the lack of the knowledge about the scalar resonances
at the simulated quark masses, 
we use a generic quartic form 
\bea
   F_S(q^2)
   & = & 
   F_S(0) \left( 1 + \frac{1}{6} \crad_S\, q^2 + c_S (q^2)^2 
                   + a_{S,3}\,(q^2)^3 + a_{S,4}\,(q^2)^4 \right)
   \label{eqn:q2:vs_q2:pff_s}
\eea
to parametrize the $q^2$ dependence of $F_S(q^2)$.
Our data are described by this form reasonably well
as in Fig.~\ref{fig:q2:pff_v+pff_s}.
The result for the scalar radius $\crad_S$
is stable against the removal of the the quartic term
as well as inclusion of higher order terms.
Such a stability is, however, not clear in the curvature $c_S$ 
due to its large statistical uncertainty.
We leave a precise determination of $c_S$ for future studies,
and only use results for $\crad_S$ in the following analysis.

\FIGURE{
   \centering 
   \includegraphics[angle=0,width=0.50\linewidth,clip]{./pff_v_vs_q2_m0025_cntrb.phys.eps}
   \vspace{-4mm}
   \caption{ 
      Contributions in the $q^2$ expansion of $F_V(q^2)\!-\!1$ 
      at $m_{ud}\!=\!0.025$. 
      Thin solid, dashed and dot-dashed lines show $O(q^2)$, $O(q^4)$ and 
      higher order contributions.
      The thick solid line is their total.
   }
   \label{fig:q2:pff_v:cntrb}
}

ChPT can provide a more unambiguous parametrization of $F_{V,S}(q^2)$.
Figure~\ref{fig:q2:pff_v:cntrb} shows 
contributions to $F_V(q^2)$ from each order $(q^2)^n$ of a Taylor expansion
of Eq.~(\ref{eqn:q2:vs_q2:pff_v}).
We observe that $O(q^6)$ and higher order contributions,
which are NNNLO and higher in ChPT, 
become a small (a few \%) correction 
below $|q^2|\!\lesssim\!(550\mbox{MeV})^2$.
Our values of $|q^2|$ are, however, outside of this region
due to the use of the simple periodic boundary condition for quark fields.
We therefore do not use a parametrization of the $q^2$ dependence
based on ChPT in this study.

We note that 
the simulated values of the pion mass squared $M_\pi^2$ 
are smaller than $(520\,\mbox{MeV})^2$.
The $O(q^6)$ contribution to $F_V(q^2)$ is small 
if $|q^2|$ is smaller than this value.
The quark mass dependence of our data of $\crad_{V,S}$
is expected to be described by NNLO ChPT.


\section{Chiral extrapolation}
 
\begin{figure}[t]
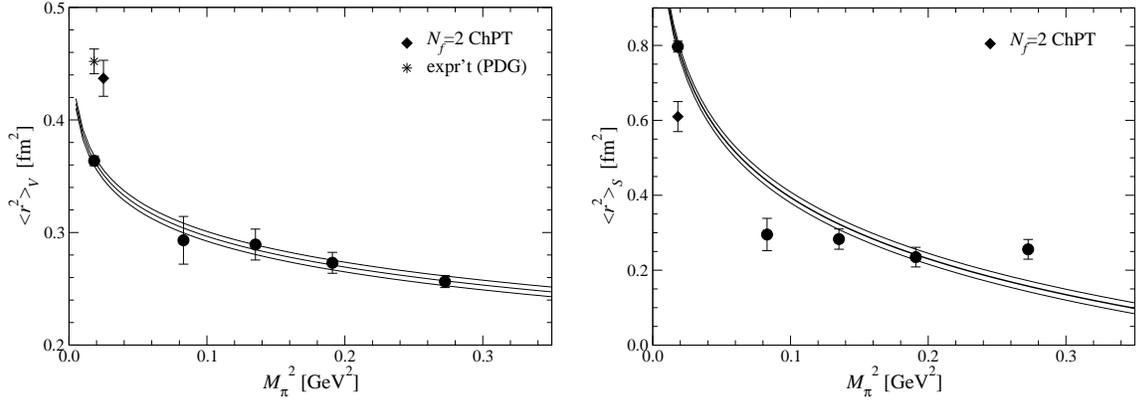

\begin{center}
   \includegraphics[angle=0,width=0.48\linewidth,clip]{./r2_v_vs_Mpi2.meas-pole+cubic.r2_v_nlo.eps}
   \hspace{3mm}
   \includegraphics[angle=0,width=0.48\linewidth,clip]{./r2_s_vs_Mpi2.norm-mom0100.quad.r2_s_nlo.eps}
   \vspace{-2mm}
   \caption{
      Chiral fit of $\crad_V$ (left panel) and $\crad_S$ (right panel)
      using NLO ChPT formulae.
      In the left panel,
      we also plot the experimental value 
      $\crad_V\!=\!0.437(16)~\mbox{fm}^2$ 
      from an analysis based on $N_f\!=\!2$ ChPT \cite{PFF_V:ChPT:NNLO} 
      (diamond) 
      and $0.452(11)~\mbox{fm}^2$ 
      quoted by Particle Data Group 
      \cite{PDG:2008} (star).
      The diamond in the right panel represents 
      $\crad_S\!=\!0.61(4)~\mbox{fm}^2$ 
      obtained from an indirect determination 
      through $\pi\pi$ scattering \cite{PFF_S:ChPT:NNLO}.
   }
   \vspace{-5mm}
   \label{fig:chiral_fit:nlo}
\end{center}
\end{figure}

We first compare our lattice data of the radii $\crad_{V,s}$
with NLO ChPT formulae \cite{PFF:ChPT:NLO}
\bea
   \crad_V
   = 
   -\frac{1}{NF^2}\left( 1 + 6 N\,l_6^r \right)
   -\frac{1}{NF^2}\ln\left[ \frac{M_\pi^2}{\mu^2} \right],
   \hspace{2mm}
   \crad_S
   = 
    \frac{1}{NF^2}\left( -\frac{13}{2} + 6 N\,l_4^r \right)
   -\frac{6}{NF^2}\ln\left[ \frac{M_\pi^2}{\mu^2} \right],
   \hspace{10mm}
   &&
   \label{eqn:chiral_fit:r2_s:nlo}
\eea
where $N\!=\!(4\pi)^2$
and $F$ is the decay constant in the chiral limit.
We fix $F$ to our estimate from our study of the pion decay constant 
\cite{Spectrum:Nf2:RG+Ovr:JLQCD}.
The renormalization scale is set to $\mu\!=\!4\pi F$.
%
The NLO fits are not quite successful 
as seen in Fig.~\ref{fig:chiral_fit:nlo}.
While our data of $\crad_V$ are fitted well with 
$\chi^2/{\rm d.o.f.}\!\sim\!0.1$,
the value extrapolated to the physical quark mass $0.364(4)~\mbox{fm}^2$ 
is significantly smaller than experiment \cite{PDG:2008}.
On the other hand, 
the NLO formula for $\crad_S$ with the enhanced chiral log
fails to reproduce our data 
and leads to large $\chi^2/{\rm d.o.f.}\!\sim\!9$.

We also note that 
NLO in ChPT is not sufficient to describe 
the quark mass dependence of the curvature $c_V$.
Although $c_V$ has a NLO term $1/(60 N F^2 M_\pi^2)$
coming from non-analytic NLO contributions to $F_V(q^2)$,
it dominates $c_V$ 
well below the physical pion mass
and fairly near the chiral limit
(see Fig.~\ref{fig:chiral_fit:nnlo:r2_c_v} below), 
where 
the non-analytic contributions to $F_V(q^2)$ are important to be consistent
with the existence of the $\gamma \to \pi\pi$ branch cut 
at $q^2\!\gtrsim\!0$.
Since $c_V$ characterizes the $O(q^4)$ dependence of $F_V(q^2)$,  
NNLO contributions are essential to describe its quark mass dependence.

\begin{figure}[b]
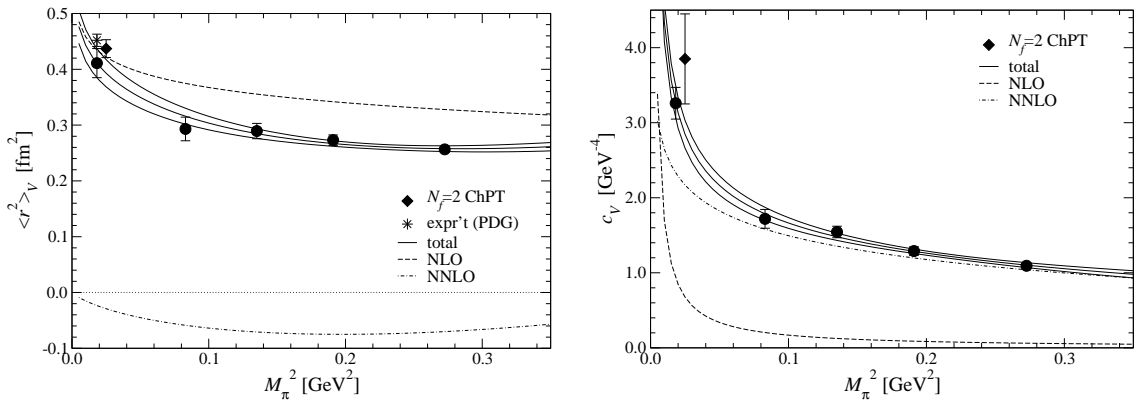

\begin{center} 
   \includegraphics[angle=0,width=0.48\linewidth,clip]{./r2_v_vs_Mpi2.meas-pole+cubic.r2_c_v.eps}
   \hspace{3mm}
   \includegraphics[angle=0,width=0.48\linewidth,clip]{./c_v_vs_Mpi2.meas-pole+cubic.r2_c_v.eps}

   \vspace{-3mm}
   \caption{
      Simultaneous chiral fit to $\crad_V$ and $c_V$
      based on two-loop formulae
      Eqs.~(\protect\ref{eqn:chiral_fit:r2_v:nnlo}) and 
      (\protect\ref{eqn:chiral_fit:c_v:nnlo}).
      We also plot a phenomenological estimate $c_V\!=\!3.85(60)$ 
      \cite{PFF_V:ChPT:NNLO}.
   }
   \label{fig:chiral_fit:nnlo:r2_c_v}
   \vspace{-5mm}
\end{center}
\end{figure}

We therefore extend our analysis to NNLO ChPT.
The NNLO contributions to the radii are 
\cite{PFF_V:ChPT:NNLO,PFF_V:ChPT:NNLO:2}
\vspace{-1mm}
\bea
   \Delta \crad_V
   & = & 
    \frac{1}{N^2F^4}
    \left( \frac{13N}{192} - \frac{181}{48} + 6 N^2 r_{V,r}^r \right)\, M_\pi^2
   +\frac{1}{N^2F^4}
    \left( \frac{19}{6} - 12 N l_{1,2}^r \right)\, 
    M_\pi^2 \ln\left[ \frac{M_\pi^2}{\mu^2} \right],
   \label{eqn:chiral_fit:r2_v:nnlo}
   \\
   \Delta \crad_S
   & = & 
    \frac{1}{N^2F^4}
    \left( - \frac{23N}{192} + \frac{869}{108} 
           + 88 N l_{1,2}^r + 80 N l_2^r + 5 N l_3^r
           - 24 N^2 l_3^r l_4^r
           + 6 N^2 r_{S,r}^r \right)\, M_\pi^2
   \nn \\
   & & 
   +\frac{1}{N^2F^4}
    \left( - \frac{323}{36} + 124 N l_{1,2}^r
                            + 130 N l_2^r \right)\, 
    M_\pi^2 \ln\left[ \frac{M_\pi^2}{\mu^2} \right] 
   -\frac{65}{3N^2F^4}
    M_\pi^2 \ln\left[ \frac{M_\pi^2}{\mu^2} \right]^2.
   \label{eqn:chiral_fit:r2_s:nnlo}
\eea   
The NNLO expression of $c_V$ is 
\bea
   c_V
   & = & 
    \frac{1}{60NF^2} \frac{1}{M_\pi^2}
   +\frac{1}{N^2F^4}
    \left( \frac{N}{720} - \frac{8429}{25920} 
                         + \frac{N}{3}l_{1,2}^r
                         + \frac{N}{6}l_6^r
                         + N^2 r_{V,c}^r \right) 
   \nn \\
   &   & 
   +\frac{1}{N^2F^4}
    \left( \frac{1}{108} + \frac{N}{3} l_{1,2}^r
                         + \frac{N}{6} l_6^r \right)\, 
    \ln\left[ \frac{M_\pi^2}{\mu^2} \right] 
   +\frac{1}{72N^2F^4}
    \ln\left[ \frac{M_\pi^2}{\mu^2} \right]^2.
   \label{eqn:chiral_fit:c_v:nnlo}
\eea
The analytic terms with $r_{\{V,S\},\{r,c\}}^r$
represent contributions from $O(p^6)$ chiral Lagrangian.
The linear combination $l_1^r\!-\!l_2^r/2$ 
appearing commonly in $\crad_V$ and $c_V$ is denoted by $l_{1,2}^r$
for simplicity.

We first carry out a simultaneous fit to $\crad_V$ and $c_V$
in terms of $M_\pi^2/(NF^2)$.
This fit has only four free parameters 
$l_6^r$, $l_{1,2}^r$, $r_{V,r}^r$ and $r_{V,c}^r$
\cite{PFF:Nf2:RG+Ovr:JLQCD},
and these can be determined with reasonable accuracy
without introducing phenomenological inputs.
As seen in Fig.~\ref{fig:chiral_fit:nnlo:r2_c_v}, 
our data are well described by the NNLO formulae
with $\chi^2/{\rm d.o.f.}\!\sim\!0.7$.
The extrapolated values of $\crad_V$ and $c_V$ 
are consistent with recent phenomenological determinations 
\cite{PFF_V:ChPT:NNLO,cV:AR,cV:GHLM}.

\begin{figure}[t]
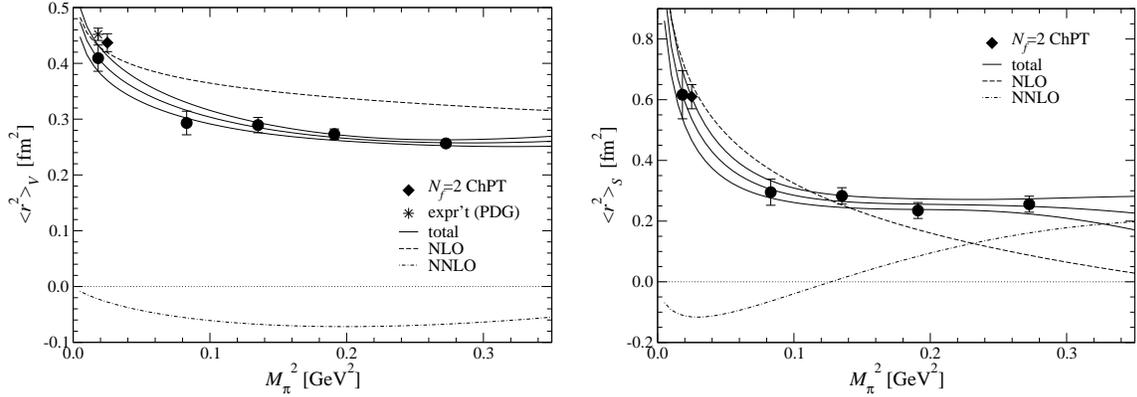

\begin{center}
   \includegraphics[angle=0,width=0.48\linewidth,clip]{./r2_v_vs_Mpi2.meas-pole+cubic.fh3-cubic.r2_v_s_c_v.vs_Mpi2.l2_l3_fixed.eps}
   \hspace{3mm}
   \includegraphics[angle=0,width=0.48\linewidth,clip]{./r2_s_vs_Mpi2.meas-pole+cubic.fh3-cubic.r2_v_s_c_v.vs_Mpi2.l2_l3_fixed.eps}
   \vspace{-3mm}
   \caption{
      Chiral extrapolation of radii $\crad_{V,S}$
      obtained from simultaneous fit to $\crad_{V,S}$ and $c_V$.
   }
   \label{fig:chiral_fit:nnlo:r2_v_s_c_v}
   \vspace{-5mm}
\end{center}
\end{figure}

The inclusion of $\crad_S$ into the simultaneous fit
introduces additional four free parameters
$l_2^r$, $l_3^r$, $l_4^r$ and $r_{S,r}^r$,
and we need to fix some of them to obtain a stable fit.
In this study, 
we use a phenomenological estimate $\bar{l}_2\!=\!4.31(11)$ 
\cite{PFF_S:ChPT:NNLO}
and a lattice estimate $\bar{l}_3\!=\!3.38(56)$ from our analysis 
of the pion mass \cite{Spectrum:Nf2:RG+Ovr:JLQCD}
\footnote{
The $\mu$--\,independent convention $\bar{l}_i$ is
defined by $l_i^r\!=\!\gamma_i(\bar{l}_i+\ln[M_\pi^2/\mu^2])/2N$
with $\gamma_1\!=\!1/3$, $\gamma_2\!=\!2/3$, 
$\gamma_3\!=\!-1/2$, $\gamma_4\!=\!2$ and $\gamma_6\!=\!-1/3$.
},
since they are determined with a reasonable accuracy 
and appear only in the NNLO terms.
As plotted in Fig.~\ref{fig:chiral_fit:nnlo:r2_v_s_c_v}, 
this fit describes our data of $\crad_S$ reasonably well
with $\chi^2/{\rm d.o.f.}\!\sim\!0.7$. 
The extrapolations of $\crad_V$ and $c_V$ are consistent 
with those in Fig.~\ref{fig:chiral_fit:nnlo:r2_c_v}.

At physical quark mass, we obtain 
\vspace{-1mm}
\bea
   \crad_V 
   = 
   0.409(23)(37)~\mbox{fm}^2,
   \hspace{2mm}
   \crad_S
   = 
   0.617(79)(66)~\mbox{fm}^2,
   \hspace{2mm}
   c_V
   = 
   3.22(17)(36)~\mbox{GeV}^{-4},
   \hspace{2mm}
\eea  
where the first and second errors are statistical and systematic, respectively.
The latter includes 
uncertainties due to the choice of the input to fix the lattice scale
and the inputs for the LECs ($F$, $l_2$ and $l_3$)
as well as 
uncertainties due to the chiral extrapolation and lattice discretization.
These results for $\crad_{V,S}$ and $c_V$ 
are consistent with phenomenological analyses.

The results for the relevant LECs are 
\bea
   &&
   \bar{l}_6
   = 
   11.9(0.7)(1.0),
   \hspace{3mm}
   \bar{l}_4
   = 
   4.09(50)(52),
   \hspace{3mm}
   \bar{l}_1-\bar{l}_2
   = 
   -2.9(0.9)(1.3),
   \\
   &&
   r_{V,r}^r = -1.0(1.0)(2.5) \times 10^{-5},
   \hspace{3mm}
   r_{V,c}^r = 4.00(17)(64) \times 10^{-5},
   \hspace{3mm}
   r_{S,r}^r = 1.74(36)(78) \times 10^{-4}.
   \hspace{10mm}
\eea
Our estimate of $\bar{l}_6$ is slightly smaller than
those from ChPT analyses:
$\bar{l}_6\!=\!16.0(0.9)$ from $F_V$ \cite{PFF_V:ChPT:NNLO}
and 15.2(0.4) from $\tau$ and $\pi$ decays \cite{l6:GPP}.
This is partly because our estimate
of $F$ \cite{Spectrum:Nf2:RG+Ovr:JLQCD} is slightly smaller 
than phenomenological estimates.
We note that $\bar{l}_4$ is consistent with our determination
$\bar{l}_4\!=\!4.12(56)$ from $F_\pi$  \cite{Spectrum:Nf2:RG+Ovr:JLQCD}
and a phenomenological estimate 4.39(22) \cite{PFF_S:ChPT:NNLO}.


\section{Conclusions} 

In this article,
we present our calculation of pion form factors 
in two-flavor lattice QCD with exact chiral symmetry,
which enables us to unambiguously compare our lattice data with two-loop ChPT.
By employing the all-to-all quark propagators,
$F_S(q^2)$ is calculated including contributions from the disconnected 
diagrams for the first time.
We observe that two-loop contributions are important 
to describe the quark mass dependence of $\crad_{V,S}$ and $c_V$
at our region of the pion mass 
$M_\pi \! \gtrsim \! 290$~MeV.
Our chiral extrapolation of $\crad_{V,S}$ and $c_V$ are consistent
with phenomenological analyses.
We also confirm that 
$F_S(q^2)$ and $F_\pi$ lead to consistent results for $l_4$.

For a more precise comparison with experiment,
we need to extend this study to three-flavor QCD. 
Such simulations are currently underway.
Another important subject is a better control of the parametrization 
of the $q^2$ dependence of $F_{V,S}(q^2)$.
To this end,
the use of the twisted boundary condition \cite{TBC}
to simulate small values of $|q^2|$, 
dispersive analysis of the $q^2$ dependence \cite{cV:AR,cV:GHLM},
and model-independent determination of the scalar resonance mass 
\cite{RoyEq:scalar:CCL} are interesting possibilities for our future studies.


\vspace{5mm}

I am grateful to my colleagues in the JLQCD and TWQCD collaborations.
I also thank Balasubramanian Ananthanarayan and Sunethra Ramanan
for a useful correspondence.
Numerical simulations are performed on Hitachi SR11000 and 
IBM System Blue Gene Solution 
at High Energy Accelerator Research Organization (KEK) 
under a support of its Large Scale Simulation Program (No.~08-05).
This work is supported in part by the Grant-in-Aid of the
Ministry of Education (No.~20105005 and 21684013).


\end{document}